\documentclass[conference]{IEEEtran}
\IEEEoverridecommandlockouts
\usepackage{cite}
\usepackage{amsmath,amssymb,amsfonts}
\usepackage{algorithmic}
\usepackage{graphicx}
\usepackage{textcomp}
\usepackage{xcolor}
\usepackage{url}
\usepackage{booktabs}

\graphicspath{ {Images/} }

\def\BibTeX{{\rm B\kern-.05em{\sc i\kern-.025em b}\kern-.08em
    T\kern-.1667em\lower.7ex\hbox{E}\kern-.125emX}}

\renewcommand{\tablename}{TABLE}
\renewcommand{\refname}{REFERENCES}

\begin{document}
\bstctlcite{IEEEexample:BSTcontrol}

\title{GNSS Receiver Operating Point Comparison Between Vector and Scalar Tracking Under High-Dynamics\\
\thanks{This work was supported by CONICET and UNLP under Grant EI004. This work has been submitted to the IEEE for possible publication. Copyright may be transferred without notice, after which this version may no longer be accessible.\\}
}

\author{
\IEEEauthorblockN{J. Fermín Llorente\textsuperscript{1,2}, Ernesto M. López\textsuperscript{1}, Javier A. Smidt\textsuperscript{1}, Pedro A. Roncagliolo\textsuperscript{1}}
\IEEEauthorblockA{\textsuperscript{1}\textit{Electronic Navigation and Telecommunications Systems (SENyT) Research Group,}\\
\textit{School of Engineering, National University of La Plata (UNLP), La Plata, Argentina.}\\
\textsuperscript{2}\textit{National Scientific and Technical Research Council (CONICET), La Plata, Argentina.}}
}

\maketitle

\begin{abstract}
This paper presents a measurement noise adaptation mechanism for Kalman filter-based carrier tracking in a navigation-aided GNSS receiver designed for launch-vehicle applications. The tracking loop is tuned for 20\,g acceleration steps, which requires a wide bandwidth that makes the loop vulnerable to noise-driven pull-out at low carrier-to-noise density ratios. To resolve this tradeoff, the measurement noise covariance, $R$, is inflated via a two-layer gate: a sigmoid envelope keyed to the signal $C/N_0$, modulated by a dynamics indicator derived from the discriminator output statistics of the Loop Bandwidth Control Algorithm. During noise-dominated periods the gate inflates~$R$, reducing the Kalman gain so the filter coasts on the navigation-aided prediction; when dynamics are detected the gate opens within one epoch, restoring the high-bandwidth tuning. The navigation aiding is formulated as a stochastic control input whose uncertainty is propagated from the navigation EKF covariance. Simulation results on a 20\,g acceleration step scenario show that the adapted vector architecture doubles the lock-retention rate relative to the standalone scalar filter at $C/N_0 = 31$\,dB-Hz. The receiver is implemented in MSGR, an open-source MATLAB GNSS software-defined receiver.
\end{abstract}


\section{Introduction}
High-dynamic Global Navigation Satellite Systems (GNSS) receivers must maintain carrier-phase lock through severe acceleration transients, such as those experienced during launch-vehicle liftoff. Engine ignition and subsequent cut-off represent some of the most stressful events during a launch campaign, where the acceleration onset is effectively modeled as a step (infinite jerk) followed by a constant acceleration. Consequently, standalone carrier tracking loops face two competing requirements: a bandwidth wide enough to absorb high-jerk dynamic transients, and sufficient noise rejection to maintain lock when the link is degraded by attenuation, multipath, or scintillation at low carrier-to-noise density ratios ($C/N_0$). The design of carrier tracking loops for high-dynamic applications has been extensively addressed in the literature, ranging from traditional digital phase-locked loops (PLLs)~\cite{article:par2012,inproceedings:Icortes2022RobustTrkSoundingRockets} to optimal tracking schemes based on Kalman Filters (KF)~\cite{inproceedings:Icortes2021AdaptiveKF,inproceedings:jfl2024,GomezFPrades2023:enhancingGNSSforRocketLaunching}. As an alternative to independent (scalar) channel tracking, vector tracking architectures are widely used to track corrupted signals by leveraging information from the highly correlated constellation~\cite{book:spilker1996a,inproceedings:Pany2006VectorDelayLock}. In these architectures, navigation estimates drive the carrier and code local replicas in the tracking channels, relying on the premise that if a sufficient number of uncorrupted signals are available, the resulting navigation solution will provide a quality of aiding that independent channels cannot achieve on their own.

Various approaches exploit this principle, including the integration of GNSS data with Inertial Navigation Systems (INS) and other sensors~\cite{GomezFPrades2023:enhancingGNSSforRocketLaunching}. GNSS-only vector tracking algorithms have also been developed~\cite{Mu2021:VTLforHighDynamicsGNSSreceiver}, with some completely replacing the traditional tracking stage with products derived from navigation filters. Loop designs capable of handling acceleration steps up to $40$\,g have been proposed~\cite{article:par2012,misc:JFL2026preprint}, successfully tracking the engine ignition dynamics of launch vehicles or sounding rockets. When tracking loops are carefully designed for these dynamics, loss of lock is rarely driven by the physical acceleration itself; rather, the primary cause of pull-out is noise at low $C/N_0$. Because the loops must be tuned to a high-bandwidth configuration to successfully track severe dynamics, they become highly susceptible to losing lock during the static pre-launch phase and during sustained acceleration after the initial tracking loop transients have settled.

To address this challenge, we incorporate algorithms that extend the inherent adaptivity of the KF architecture. The first is the Loop Bandwidth Control Algorithm (LBCA)~\cite{inproceedings:Icortes2021AdaptiveKF}, which adaptively modifies the loop bandwidth based on discriminator output statistics. This allows the receiver to narrow the bandwidth during low-dynamic, high-noise conditions and widen it during high-dynamic events. We combine this with a navigation-aided vector tracking philosophy, where navigation estimates provide a more reliable, information-rich baseline for the signal parameters. This aiding has been proposed in \cite{article:gao2024VTRwitAdaptiveQplusDelta}, although only done with frequency aiding and not developed as a KF with stochastic control input as it is framed in this paper. Because these estimates are derived from the aggregate of multiple signals, the tracking stage remains robust even if an individual channel experiences severe fading or outage. By fusing these approaches, the tracking loops lean on continuously available navigation estimates while the LBCA adaptively tunes the bandwidth. During the peak stress of the acceleration step, the filter maintains its nominal dynamic tuning; however, before this event and once the transients settle, the bandwidth is dynamically reduced to maximize noise rejection and minimize pull-out probability.

To validate these algorithms, we require a MATLAB-based vector GNSS software receiver with access to the tracking loop internals at the filter-coefficient level. A comprehensive survey of available GNSS SDRs is provided in~\cite{inproceedings:akosEtAl2022GNSSsdrHistoryAndRecopilation}. However, the existing MATLAB implementations are either designed for general-purpose positioning applications or are available only through legacy, unmaintained repositories. We therefore selected the CU-SDR Collection~\cite{inproceedings:Bernabeu2022CollectionSDRsGNSS} as a lightweight, well-documented GPS~L1 baseline and adapted it to support the vector tracking architecture described in this work, providing a reproducible testbed for the results presented in this paper.
% To validate these algorithms, we adapted an open-source, MATLAB-based GNSS software receiver to support a vector tracking architecture. While a MATLAB-based open-source vector receiver was previously published~\cite{article:gpsSol2019OpenSourceMatlabVTReceiver}, its source code is no longer publicly accessible at the time of writing. Other open-source legacy alternatives exist~\cite{inproceedings:ZhaoAkos2011OpenSourceVTLreceiver}, although they. Our adaptation effectively restores and expands this capability, providing a versatile testbed for evaluating advanced vector tracking loops and ensuring the reproducibility of the results presented in this paper.
%The remainder of this paper is organized as follows. Section~\ref{sec:VTL} describes the vector tracking architecture with navigation aiding and the LBCA-gated $R$~adaptation. Section~\ref{sec:MSGR} presents the MSGR platform and test configuration. Section~\ref{sec:results} reports simulation results, and Section~\ref{sec:conclusions} concludes.
\section{Vector tracking architecture}\label{sec:VTL}
Vector Tracking Loops (VTL) couple the tracking-channel loops with the navigation stage by obtaining estimates for code phase, carrier and code frequency from the navigation filter. Different architectures exist, depending on the extent of coupling. A federated version where the feedback is done at a different rate than the update rate of the tracking loops was proposed in~\cite{inproceedings:Pany2006VectorDelayLock}. However, when it is possible to run navigation and tracking at the same rates a centralized approach can be implemented, carrier and code replica generators are driven by the navigation stage estimates. This latter architecture was proposed in~\cite{inproceedings:Siebert2022centralizedMultipathRejection} and its performance was compared with the federated approach in~\cite{inproceedings:Siebert2024comparisonNAVITEC}. In this work we are going to close the loop at the rate imposed by the tracking stage, each $T$ seconds both tracking and navigation products are computed, following a philosophy similar to~\cite{inproceedings:Siebert2022centralizedMultipathRejection}. The primary difference here is that we still rely on KF-based carrier tracking loops designed specifically to handle high-dynamics such as acceleration steps~\cite{misc:JFL2026preprint}. These loops have optimal performance for this dynamic profile, with a tracking threshold of $\sim 31$\,dB-Hz for $T = 5$\,ms. To extend the operative regime of our receiver, aiding the tracking loops with kinematic estimates from navigation allows the loops to adaptively reduce their bandwidth while maintaining lock under lower $C/N_0$ conditions than a scalar architecture could achieve. This is a fundamental gain obtained with vector architectures. We formulate this aiding in a stochastic control framework by modifying the time update equation of the state-space model through the addition of a control input with uncertainty.

% Since tracking loops operating at such a low CN0 (and high bandwidth) are in danger of directly pulling out even in a static operation, adaptively lowering the bandwidth based on an estimate of the current dynamics regime ensures that the loops survive the static pre-launch phase. The bandwidth is then effectively incremented when the dynamic event occurs, returning to the design value known to correctly handle the high-jerk profile. However, because this lowered bandwidth would otherwise render the loops incapable of tracking sudden signal dynamics, an aiding signal from the navigation stage is added as a control input to bridge this vulnerability.

\subsection{Navigation-Aided mKF with Adaptive R}\label{ssec:VTL_arch}
% --- Navigation EKF ---
The navigation stage seeds an Extended Kalman Filter (EKF) with an initial least-squares position fix obtained after decoding the navigation data. The EKF state vector comprises position, velocity, and acceleration in ECEF coordinates, together with receiver clock bias and drift:
\begin{equation}\label{eq:navEKF_state}
  \mathbf{x}_k^{\text{nav}} =
  \begin{bmatrix}
    \mathbf{p}_k^T & \mathbf{v}_k^T & \mathbf{a}_k^T & t_b & t_d
  \end{bmatrix}^T \in \mathbb{R}^{11},
\end{equation}
with state transition $\mathbf{x}_{k+1}^{\text{nav}} = \mathbf{F}_{\text{nav}}\,\mathbf{x}_k^{\text{nav}} + \mathbf{w}_k$, where~\cite{book:brown4ed2012}\cite{book:kaplan2006}
\begin{equation}\label{eq:navEKF_F}
  \mathbf{F}_{\text{nav}} =
  \begin{bmatrix}
    \boldsymbol{\Phi} \otimes \mathbf{I}_3
      & \mathbf{0}_{9\times 2} \\[2pt]
    \mathbf{0}_{2\times 9}
      & \mathbf{F}_{\text{clk}}
  \end{bmatrix},
\end{equation}
\begin{equation}\label{eq:navEKF_Phi}
  \boldsymbol{\Phi} =
  \begin{bmatrix}
    1 & T & T^2/2 \\ 0 & 1 & T \\ 0 & 0 & 1
  \end{bmatrix}, \qquad
  \mathbf{F}_{\text{clk}} =
  \begin{bmatrix}
    1 & T \\ 0 & 1
  \end{bmatrix},
\end{equation}
$T$ is the filter update period, and $\mathbf{w}_k \sim \mathcal{N}(\mathbf{0},\mathbf{Q}_{\text{nav}})$. The Kronecker product $\boldsymbol{\Phi}\otimes\mathbf{I}_3$ applies the scalar kinematic coupling of~$\boldsymbol{\Phi}$ identically to each ECEF axis.

% --- Per-channel tracking with aiding ---
The per-channel carrier tracking loop uses the modified Kalman filter (mKF) of~\cite{misc:JFL2026preprint}, which employs a delay-aware measurement model, extended here to a vector architecture by adding a navigation aiding term to the predict step. For each channel, the navigation EKF's velocity and acceleration estimates are projected onto the line-of-sight (LOS) unit vector $\mathbf{e}_{\text{LOS}}$ to obtain frequency and frequency-rate aiding:
\begin{equation}\label{eq:u_nav}
  \mathbf{u}_k^{\text{nav}} =
  \begin{bmatrix}
    0 \\[2pt]
    \mathbf{h}_{\text{LOS}}^T \mathbf{v}_k \\[2pt]
    \mathbf{h}_{\text{LOS}}^T \mathbf{a}_k
  \end{bmatrix}, \qquad
  \mathbf{h}_{\text{LOS}} = -\frac{\mathbf{e}_{\text{LOS}}}{\lambda},
\end{equation}
where $\lambda$ is the GPS L1 carrier wavelength. The projection vector $\mathbf{h}_{\text{LOS}}$ converts velocity~[m/s] and acceleration~[m/s$^2$] to carrier frequency~[Hz] and frequency rate~[Hz/s] along the satellite LOS. The tracking-loop state-space model becomes
\begin{equation}\label{eq:stateTransitionModelKFwControl}
  \begin{cases}
    \hat{\mathbf{x}}_{k+1}^{-} = \mathbf{F}\,\hat{\mathbf{x}}_k^{+}
      + \mathbf{G}\,\mathbf{u}_k^{\text{nav}} \\[4pt]
    y_k = \mathbf{H}\,\mathbf{x}_k + v_k
  \end{cases}
\end{equation}
with
\begin{equation}\label{eq:FG}
  \mathbf{F}\!=\!
  \begin{bmatrix}
    1 & T & \frac{T^2}{2} \\[2pt]
    0 & 1 & T \\[2pt]
    0 & 0 & 1
  \end{bmatrix}\!,\;
  \mathbf{G}\!=\!
  \begin{bmatrix}
    0 & T & \frac{T^2}{2} \\[2pt]
    0 & 0 & T \\[2pt]
    0 & 0 & 0
  \end{bmatrix}\!,
\end{equation}
\begin{equation}\label{eq:H_delay}
  \mathbf{H} =
  \begin{bmatrix}
    1 & -T & T^2\!/2
  \end{bmatrix}\!,
\end{equation}
where $\mathbf{H}$ is the delay-aware measurement matrix and $S_j$ the jerk power spectral density of~$\mathbf{Q}$, both from~\cite{misc:JFL2026preprint}. In the vector configuration, frequency and frequency-rate tracking states represent residual errors after the navigation projection is applied in the time update; in the scalar (standalone) configuration, $\mathbf{u}_k^{\text{nav}} = \mathbf{0}$ and the states are absolute estimates.

% --- Covariance augmentation ---
To maintain a calibrated covariance under aiding, the predict-step error covariance is augmented with the navigation input uncertainty
\begin{equation}\label{eq:Pminus_aug}
  \mathbf{P}_k^{-} = \mathbf{F}\,\mathbf{P}_{k-1}^{+}\mathbf{F}^T
    + \mathbf{Q}
    + \mathbf{G}\,\boldsymbol{\Sigma}_{\mathbf{u}}^{\text{nav}}
      \mathbf{G}^T \ ,
\end{equation}
where
\begin{equation}\label{eq:Sigma_u}
  \boldsymbol{\Sigma}_{\mathbf{u}}^{\text{nav}} =
  \begin{bmatrix}
    0 & \mathbf{0}_{1\times 2} \\[2pt]
    \mathbf{0}_{2\times 1} &
      \boldsymbol{\Sigma}_{\text{aid}}
  \end{bmatrix} \ ,
\end{equation}
\begin{equation}\label{eq:Sigma_aid}
  \boldsymbol{\Sigma}_{\text{aid}} =
  \mathbf{H}_{\text{proj}}\,
  \mathbf{P}_{\text{aid}}^{\text{nav}}\,
  \mathbf{H}_{\text{proj}}^T
  +
  \begin{bmatrix}
    \sigma_{t_d}^2\!/\lambda^2 & 0 \\[2pt]
    0 & 0
  \end{bmatrix} \ .
\end{equation}
Here $\mathbf{P}_{\text{aid}}^{\text{nav}}$ is the $6\!\times\!6$ velocity--acceleration covariance block extracted from the navigation EKF
\begin{equation}\label{eq:P_aid}
  \mathbf{P}_{\text{aid}}^{\text{nav}} =
  \begin{bmatrix}
    \text{cov}(\mathbf{v},\mathbf{v}) &
      \text{cov}(\mathbf{v},\mathbf{a}) \\[2pt]
    \text{cov}(\mathbf{a},\mathbf{v}) &
      \text{cov}(\mathbf{a},\mathbf{a})
  \end{bmatrix}\! ,
%   \qquad
  \mathbf{H}_{\text{proj}} =
  \begin{bmatrix}
    \mathbf{h}_{\text{LOS}}^T & \mathbf{0}_{1\times 3} \\[2pt]
    \mathbf{0}_{1\times 3} & \mathbf{h}_{\text{LOS}}^T
  \end{bmatrix}\!,
\end{equation}
and $\sigma_{t_d}^2$ is the clock-drift variance from the navigation covariance, projected to carrier frequency via the factor~$1/\lambda^2$ (consistent with the clock-drift state in~m/s). This augmentation is the standard covariance treatment for a Kalman filter with stochastic control input~\cite{book:barShalom2004}: it ensures that~$\mathbf{P}^{-}$ honestly reflects the prediction uncertainty from imperfect aiding, and it modifies the steady-state gain to correctly balance the navigation prior against the discriminator measurement.

% --- Effect of aiding on the gain ---
The mKF is tuned via~$S_j$ to withstand 20\,g acceleration steps. Since~$\mathbf{K}$ depends on~$\mathbf{P}^{-}$ and~$R$ but not on~$\hat{\mathbf{x}}_k^{-}$, the injection~$\mathbf{G}\hat{\mathbf{u}}_k$ improves the prediction without changing the gain, while the augmentation in Eq.~\eqref{eq:Pminus_aug} only adds the aiding uncertainty on top of the 20\,g process noise in~$\mathbf{P}^{-}$. Once the aiding has settled, the tracking states follow only mild residual dynamics, yet the filter retains the bandwidth sized for the 20\,g transient. At low~$C/N_0$ this excess bandwidth admits noise that drives pull-out, so the threshold of the dynamics-optimized tuning persists under aiding alone, motivating the adaptive~$R$ developed next.

\subsection{LBCA-gated R adaptation}\label{ssec:lbca_gated_R}
The measurement noise variance $R_k$ is conventionally set to the discriminator's output approximated variance, \mbox{$R_0 = 1/(2TC/N_0)$}~\cite{book:spilker1996a}. Near the tracking threshold, the small-phase-error and Gaussian-noise assumptions underlying this model break down, and pull-out becomes noise-driven. In the navigation-aided architecture, inflating $R$ drives $\mathbf{K}\to\mathbf{0}$, letting the filter coast on the $\mathbf{G}\hat{\mathbf{u}}_k$ prediction. However, a pure $C/N_0$-driven inflation would also de-weight the discriminator during dynamics transients when the aiding is transiently stale. We therefore gate the inflation using the LBCA discriminator statistics~\cite{inproceedings:Icortes2021AdaptiveKF}, which distinguish noise-dominated from dynamics-dominated regimes at the per-epoch rate.

\subsubsection{Discriminator statistics}

Three metrics derived from the phase discriminator output $u_k$ are maintained via an exponential moving average (EMA) with forgetting factor $\gamma$:
\begin{align}
  \mu_{|u|}[k] &= \gamma\,\mu_{|u|}[k\!-\!1]
    + (1-\gamma)\,|u_k| \label{eq:ema_abs}\\
  \mu_u[k] &= \gamma\,\mu_u[k\!-\!1]
    + (1-\gamma)\,u_k \\
  \mu_{u^2}[k] &= \gamma\,\mu_{u^2}[k\!-\!1]
    + (1-\gamma)\,u_k^2
\end{align}
from which the smoothed standard deviation \mbox{$\sigma_u = \sqrt{\max(0,\;\mu_{u^2} - \mu_u^2)}$} and the normalized dynamics and noise indicators are computed as~\cite{inproceedings:Icortes2021AdaptiveKF}
\begin{equation}\label{eq:DN}
  D_k = \frac{\mu_{|u|}}{\mu_{|u|} + \sigma_u}\,,\qquad
  N_k = \frac{\sigma_u}{\mu_{|u|} + \sigma_u}\,\ .
\end{equation}

When the discriminator output is noise-dominated (zero mean, spread dominates), $D_k$ is low. When a consistent bias is present, indicating a dynamics-driven phase ramp that the loop has not yet absorbed, $D_k$ rises towards unity. It is evident from Eq.~\eqref{eq:DN} that both metrics sum one.

\subsubsection{Two-layer R gate}

The effective measurement noise variance combines a $C/N_0$-dependent sigmoid envelope with the LBCA dynamics gate
\begin{equation}\label{eq:R_adapt}
  R_{\mathrm{raw},k} = R_{0,k}\!\left(
    1 + \alpha\;S_k\;g_k
  \right) \ .
\end{equation}
where $S_k, g_k \in [0,1]$ are the $C/N_0$ envelope and the dynamics gate defined below, and the inflation gain $\alpha \geq 0$ bounds $R_{\mathrm{raw},k}/R_{0,k}$ to $[1,\, 1+\alpha]$, the upper bound being reached only when both gates are fully active.

The sigmoid envelope determines when inflation is appropriate
\begin{equation}\label{eq:sigmoid}
  S_k = \frac{1}{1 + e^{-\beta\,(C/N_{0,\mathrm{th}} - C/N_{0,k})}}
\end{equation}
where $C/N_{0,\mathrm{th}}$ is the standalone tracking threshold (in~\mbox{dB-Hz}), $\beta$~(dB$^{-1}$) controls the transition steepness, and $C/N_{0,k}$ is a smoothed $C/N_0$ estimate (decoupled from per-epoch estimation noise). At $C/N_0 \gg C/N_{0,\mathrm{th}}$, $S_k\!\approx\!0$ and the adaptation is dormant. Below the threshold, $S_k\!\to\!1$.

The dynamics gate determines whether to inflate at the current epoch:
\begin{equation}\label{eq:gate}
  g_k = \max\!\left(0,\;1 - \frac{D_k}{D_{\mathrm{th}}}\right)
\end{equation}
When $D_k < D_{\mathrm{th}}$ (noise dominates), $g_k > 0$ and $R$ is inflated. When $D_k \geq D_{\mathrm{th}}$ (dynamics detected), $g_k = 0$ and $R$ returns to the theoretical optimal baseline, allowing the Kalman gain and loop bandwidth to recover.

\subsubsection{Asymmetric R smoother}

To prevent false inflation from single noisy epochs while ensuring instant response to dynamics, $R_{\mathrm{raw},k}$ is also smoothed asymmetrically as follows
\begin{equation}\label{eq:asym_smoother}
  R_{\mathrm{eff},k} = \gamma_R\,R_{\mathrm{eff},k-1}
    + (1 - \gamma_R)\,R_{\mathrm{raw},k}
\end{equation}
where the smoothing factor switches between two values:
\begin{equation}
  \gamma_R = \begin{cases}
    \gamma_{\mathrm{fall}} & \text{if }
      R_{\mathrm{raw},k} < R_{\mathrm{eff},k-1}
      \quad\text{(dynamics onset)}\\[2pt]
    \gamma_{\mathrm{rise}} & \text{otherwise}
      \quad\text{(noise onset)}
  \end{cases}
\end{equation}
With $\gamma_{\mathrm{fall}} = 0$ (one-epoch step-down) and $\gamma_{\mathrm{rise}} = 0.9$ (gradual inflation), the mechanism restores the baseline tuning within a single epoch when dynamics are detected, with the loop bandwidth then recovering at the rate of the filter's covariance dynamics, while requiring sustained noise dominance before inflating~$R$.

The three stages operate in cascade: the discriminator statistics \eqref{eq:ema_abs}--\eqref{eq:DN} detect the regime, the two-layer gate~\eqref{eq:R_adapt} computes the target inflation, and the asymmetric smoother~(\ref{eq:asym_smoother}) shapes the transient response. The combined mechanism is self-regulating: if the inflation causes the loop to miss a dynamics event (growing $|u_k|$ bias), $D_k$ rises, the gate closes, $R$ drops, $\mathbf{K}$ increases, and the discriminator corrects the loop, without external intervention or an explicit state machine.

\section{MSGR Platform and Experimental Setup}\label{sec:MSGR}
\subsection{MSGR overview}
The MATLAB SENyT GNSS Receiver (MSGR)~\cite{misc:MSGRrepo} is an open-source GPS~L1~C/A software-defined receiver developed at SENyT, Universidad Nacional de La Plata, Argentina. MSGR was forked from the CU-SDR Collection~\cite{inproceedings:Bernabeu2022CollectionSDRsGNSS} and substantially extended for research on KF-based tracking and vector architectures in high-dynamics applications.
 
%The principal additions over the CU-SDR baseline are:
% \begin{itemize}
%   \item a delay-aware modified Kalman filter (mKF) carrier tracking loop~\cite{misc:JFL2026preprint}, replacing the original PLL/FLL structure;
%   \item a 11-state navigation EKF with configurable process noise and update rate (Section~\ref{ssec:VTL_arch});
%   \item vector tracking support with per-channel navigation aiding injection and automatic scalar/vector mode switching;
%   \item the LBCA-gated $R$ adaptation described in Section~\ref{ssec:lbca_gated_R};
%   \item configurable coherent integration time with a histogram-based bit synchronization algorithm;
%   \item a Skydel/GSG-8 complex-baseband data interface for controlled simulation campaigns.
% \end{itemize}
%
The principal additions over the CU-SDR baseline are: an mKF carrier tracking loop~\cite{misc:JFL2026preprint} (Section~\ref{ssec:VTL_arch}) replacing the original PLL structure; an \mbox{11-state} navigation EKF with configurable process noise and update rate (Section~\ref{ssec:VTL_arch}); vector tracking support with per-channel navigation aiding injection and automatic scalar/vector mode switching; the LBCA-gated $R$~adaptation described in Section~\ref{ssec:lbca_gated_R}; configurable coherent integration time with a histogram-based bit synchronization algorithm; and a Skydel/GSG-8~\cite{misc:safranGSGmanual} complex-baseband data interface for controlled simulation campaigns.

The acquisition stage, correlator structure, and measurement generation core are retained from the CU-SDR Collection with minor modifications. The receiver can be configured to operate in scalar (standalone~mKF) or vector (navigation-aided~mKF) mode through a unified processing pipeline, enabling direct comparison of both architectures on the same dataset.

\subsection{Test configuration}
The datasets were generated with a Safran GSG-8 GNSS signal generator running the Skydel simulation engine~\cite{misc:safranGSGmanual}. GPS~L1~C/A signals are recorded as complex baseband samples at $f_s = 12.5$\,MHz with 16-bit quantization and processed offline with~MSGR.
 
A constellation of 6~GPS satellites is tracked, with all channels at nominal $C/N_0 \approx 40$\,dB-Hz except PRN\,15, whose power is reduced to evaluate tracking near the standalone threshold. The simulated trajectory models a launch vehicle with a static pre-launch phase followed, at $t = 40$\,s, by a vertical acceleration step (held constant thereafter) scaled so that its projection onto the LOS of PRN~15 is 20\,$g$ ($\approx 196$\,m/s$^2$). This delay allows sufficient time for navigation message decoding, initial position fix, EKF convergence, and vector tracking mode initialization. Although~\cite{misc:JFL2026preprint} uses $T = 5$\,ms, here $T = 10$\,ms only shortens offline runtimes, navigation at the tracking rate is demanding but feasible on modern embedded platforms.

To demonstrate the LBCA-gated $R$ adaptation with a limited number of noise realizations, the test $C/N_0$ for PRN\,15 is set to 31\,dB-Hz, an operating point where the standalone pull-out probability is well above 0.1 and the scalar mKF fails in the majority of realizations, so the effect of the $R$~adaptation is visible without large Monte Carlo campaigns.
 
The operating point is processed across 18~independent noise realizations while keeping the trajectory and constellation geometry identical. For each realization, the receiver is run in two configurations: scalar mode (standalone mKF, $\mathbf{u}_k^{\text{nav}} = \mathbf{0}$, nominal~$R$) and vector mode (navigation-aided mKF with LBCA-gated $R$~adaptation). The LBCA parameters are: discriminator EMA factor $\gamma = 0.9$, sigmoid steepness $\beta = 1.0$\,dB$^{-1}$ centered at $C/N_{0,\text{th}} = 31$\,dB-Hz, maximum inflation ratio $\alpha = 50$, dynamics threshold $D_{\text{th}} = 0.5$, and asymmetric smoother factors $\gamma_{\text{rise}} = 0.9$, $\gamma_{\text{fall}} = 0$ (one-epoch restoration at dynamics onset).

\section{RESULTS}\label{sec:results}

\subsection{Baseline Comparison at 35\,dB-Hz}

\begin{figure}[]
    \centering
    \includegraphics[width=0.49\textwidth]{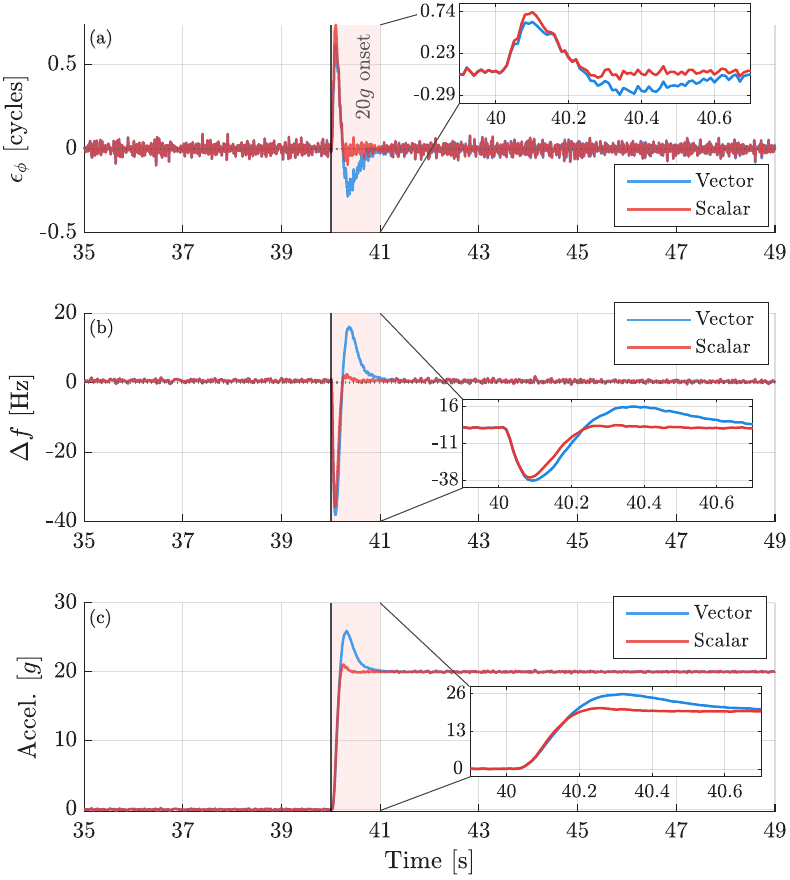}
    \caption{Comparison of signal parameter estimates used for NCO configuration: (a)~phase discriminator output, (b)~carrier frequency, and (c)~line-of-sight acceleration. Standalone scalar and vector aided with LBCA-gated $R$ adaptation at $C/N_0 = 35$\,dB-Hz, $T = 10$\,ms.}
    \label{fig:trkComparison_35dB-Hz}
\end{figure}

Figure~\ref{fig:trkComparison_35dB-Hz} compares the standalone scalar mKF and the navigation-aided vector mKF with LBCA-gated $R$~adaptation at $C/N_0 = 35$\,dB-Hz, well above the standalone tracking threshold. Both architectures maintain lock through the 20\,g acceleration step. The phase discriminator output (Fig~\ref{fig:trkComparison_35dB-Hz}a) shows comparable noise levels, confirming that the $R$~adaptation does not degrade tracking performance when the signal is strong, the sigmoid factor $S_k$ is near zero at this $C/N_0$ and the LBCA remains dormant. The transient response of the vector architecture shows a slightly different overshoot profile due to the navigation EKF's convergence transient, visible in all three states. It is important to note that the magnitudes in this plot correspond to the estimates fed to the NCOs, thus consist of the mKF plus the navigation aiding estimates.

% Figure — Tracking Comparison	- Export at width 13.319(cm)	and heigth 14.526(cm)

\subsection{Navigation-aided tracking at 31\,dB-Hz}

At $C/N_0 = 31$\,dB-Hz, the standalone scalar mKF loses lock due to noise in the majority of realizations. The wide bandwidth required by the 20\,g tuning ($S_j = 46$\,dB) admits more thermal noise than the loop can tolerate at this signal level, even during the static pre-launch phase.

\begin{figure}[]
    \centering
    \includegraphics[width=0.49\textwidth]{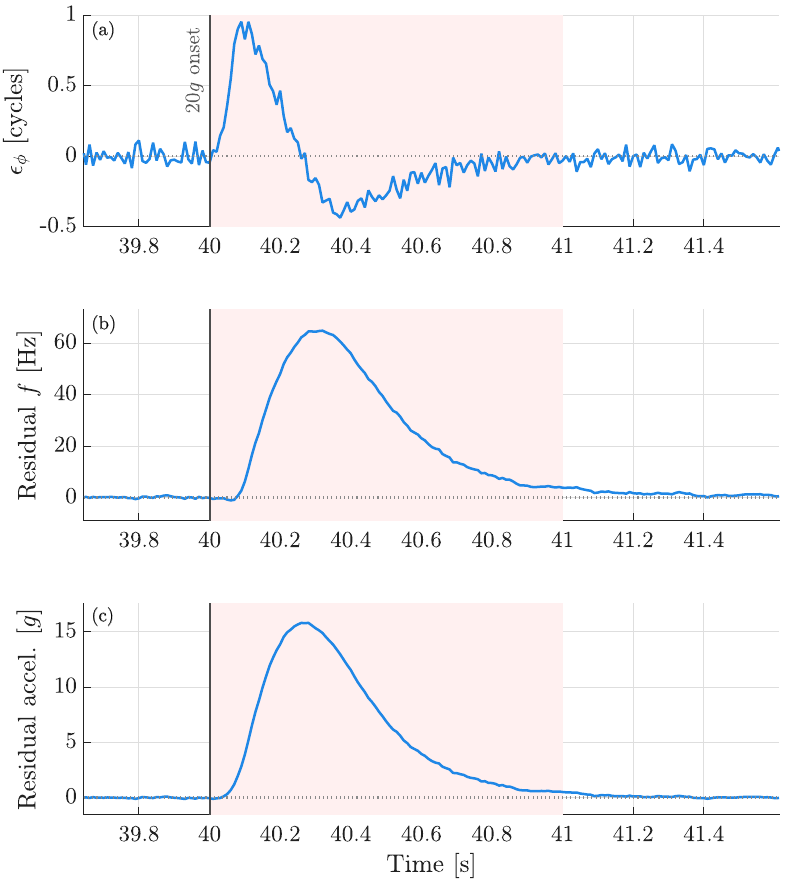}
    \caption{Tracking KF state estimates using vector aided with LBCA-gated $R$ adaptation at $C/N_0 = 31$\,dB-Hz: (a)~phase discriminator output, (b)~frequency residual, and (c)~acceleration residual.}
    \label{fig:trkVectorAided_31dB-Hz}
\end{figure}

Figure~\ref{fig:trkVectorAided_31dB-Hz} shows the vector mKF with LBCA-gated $R$~adaptation maintaining lock through the acceleration step at 31\,dB-Hz. The three panels show the tracking-loop residuals after navigation aiding, zoomed around the step onset. The phase discriminator output (Fig.~\ref{fig:trkVectorAided_31dB-Hz}a) shows a transient excursion approaching $\pm 1$~cycle at the step, settling within approximately 0.8\,s. The tracking loop now estimates the frequency residual (Fig.~\ref{fig:trkVectorAided_31dB-Hz}b) since the mean trajectory of the frequency is the aiding injection. Once the navigation EKF converges, the frequency residual stabilizes and effectively goes to zero. The acceleration residual (Fig.~\ref{fig:trkVectorAided_31dB-Hz}c) peaks at ${\sim}\,15$\,g (the portion of the 20\,g step not yet estimated by navigation) and decays over approximately 1\,s. After convergence, all three residuals return to their pre-step noise floor, confirming that the bandwidth adaptation is effective.

% Figure — LBCA Tracking Detail	- Export at width 13.266(cm)	and height 14.552(cm)

\begin{figure}[]
    \centering
    \includegraphics[width=0.49\textwidth]{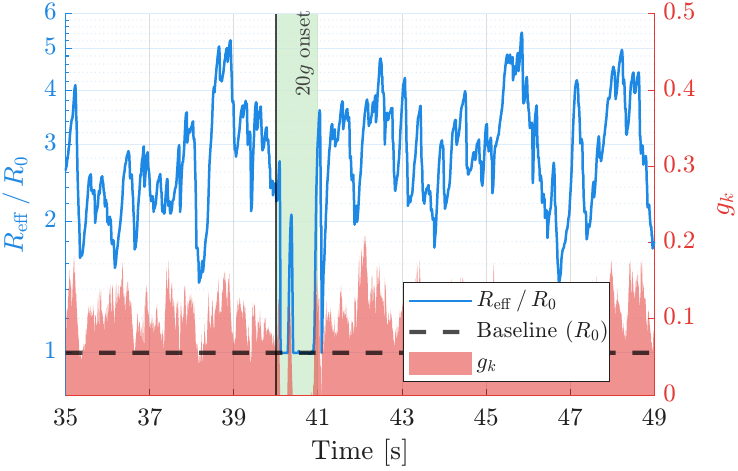}
    \caption{LBCA mechanism triggering the adaptive bandwidth control at $C/N_0 = 31$\,dB-Hz and in the aiding vector mode.}
    \label{fig:LBCAmechanism_31dB-Hz}
\end{figure}

Figure~\ref{fig:LBCAmechanism_31dB-Hz} shows the LBCA mechanism operating during the same scenario. The measurement noise ratio $R_{\text{eff}}/R_0$ (left axis) fluctuates between 1.5 and 5 during the noise-dominated pre-step phase, driven by the dynamics gate $g_k$ (right axis, shaded). At the acceleration step onset (shaded region), $g_k$ drops sharply as the $D_k$ statistic detects the dynamics transient, causing $R_{\text{eff}}$ to fall toward the baseline $R_0$ (the actual covariance of the phase error model). This restores the full Kalman gain and loop bandwidth, allowing the discriminator to guide the loop through the transient. After approximately 1\,s, the dynamics settle, $g_k$ recovers, and $R_{\text{eff}}$ rises again to suppress noise. The asymmetric smoother ensures one-epoch response on the falling edge (dynamics onset) and gradual recovery on the rising edge (return to noise dominance).

 % Figure — LBCA Summary Compact - Export at width 13.161(cm) and heigth	8.334(cm)

\subsection{Performance comparison}

Across 18 independent noise realizations at \mbox{$C/N_0 = 31$\,dB-Hz}, the standalone scalar mKF maintained lock in 3 of 18 realizations (17\%), while the LBCA-gated vector architecture maintained lock in 6 of 18 (33\%). The pull-out events that were avoided with the LBCA-gated vector architecture correspond to pull-outs at static or constant acceleration (with tracking loops settled) conditions. Both architectures operate well below the nominal tracking threshold at this $C/N_0$, and the high failure rates reflect the elevated pull-out probability inherent to the dynamics-optimized tuning. The doubling of the survival rate is consistent with the moderate effective inflation ($R_{\mathrm{eff}}/R_0 \approx 2$--$5$, Fig.~\ref{fig:LBCAmechanism_31dB-Hz}): although $\alpha$ sets its upper bound, it is further scaled by $S_k$ and $g_k$, which stay well below unity at this noise level, where the LBCA dynamics gate approaches its detection limit. With the sigmoid centered at the actual standalone threshold (${\sim}\,32.5$\,dB-Hz, calculated empirically as in~\cite{misc:JFL2026preprint}), the mechanism operates in the same regime relative to the threshold and a comparable improvement is expected.

\subsection{Discussion}

The demonstrated improvement is modest in absolute terms but addresses a specific operational vulnerability: the noise--dynamics tradeoff inherent in fixed-tuning tracking loops designed for worst-case dynamics. Two remarks frame the scope of these results. First, in open-sky launch-vehicle scenarios $C/N_0$ typically lies in the 35--45\,dB-Hz range; the 31\,dB-Hz point is not nominal but emulates transient degradations such as antenna-pattern nulls, plume attenuation, or scintillation, which the receiver should flag, yet retaining lock through them remains highly valuable. Second, such degradations often affect several satellites in the same sky region simultaneously, so adding constellations (e.g., Galileo) does not remove the need for low-$C/N_0$ robustness; rather, it is complementary, as more measurements strengthen the navigation aiding on which the $R$~inflation relies. The LBCA-gated $R$~adaptation does not require re-tuning the mKF for each $C/N_0$ level, the process noise $S_j$ handles the dynamics, while the $R$~adaptation handles the noise. The mechanism's effectiveness is bounded by the LBCA dynamics indicator's ability to distinguish dynamics from noise: as $C/N_0$ decreases, the discriminator noise floor rises and the $D_k$ baseline approaches $D_{\text{th}}$, reducing the detection margin. Inertial aiding~\cite{GomezFPrades2023:enhancingGNSSforRocketLaunching} is expected to outperform this GNSS-only approach at higher integration cost. However, the $R$~adaptation is agnostic to the aiding source, and an INS-coupled filter would simply provide $\mathbf{u}_k^{\text{nav}}$ and $\boldsymbol{\Sigma}_{\mathbf{u}}^{\text{nav}}$. A rigorous statistical characterization at $C/N_0$ values near the standalone threshold, where the pull-out probability is lower and the $R$~adaptation has more margin, would require Monte Carlo campaigns with substantially more realizations and is left for future work, potentially leveraging the pull-out probability framework of~\cite{misc:JFL2026preprint} extended to the adaptive-$R$ architecture.

% =====================================================================

\section{Conclusions}\label{sec:conclusions}

This paper presented a measurement noise adaptation mechanism for navigation-aided KF-based carrier tracking in a launch-vehicle GNSS receiver. The LBCA-gated $R$~adaptation uses discriminator statistics to resolve the noise--dynamics tradeoff in tracking loops tuned for high-dynamics applications: during noise-dominated periods, $R$ is inflated to reduce the Kalman gain and suppress noise-driven pull-outs; when dynamics are detected, $R$ is restored within one epoch to preserve the loop's dynamic capability. The mechanism is self-regulating and requires no explicit state machine or dynamics profile knowledge.

The navigation aiding, formulated as a stochastic control input with uncertainty derived from the navigation EKF covariance, provides the accurate state prediction that makes the $R$~inflation viable. Without it, reducing the Kalman gain would cause the loop to drift. The LBCA dynamics gate ensures that the inflation is suspended during transients where the aiding is transiently stale.

Simulation results on a 20\,$g$ LOS acceleration step show that the LBCA-gated vector architecture approximately doubles the lock-retention rate (from 17\% to 33\%) relative to the standalone mKF at $C/N_0 = 31$\,dB-Hz. The open-source MSGR receiver platform used for this evaluation is publicly available~\cite{misc:MSGRrepo}.

Future work includes Monte Carlo characterization near the standalone tracking threshold for other acceleration profiles and integration times, analytical prediction of the adapted threshold via the pull-out probability framework of~\cite{misc:JFL2026preprint}, extension to signal blackout and coast/reacquisition scenarios, and validation with hardware-in-the-loop RF signals.

\section*{Acknowledgement}
The authors would like to thank \textit{Safran Electronics \& Defense} for providing the Skydel GNSS Simulation software, as part of the Minerva Academic Partnership Program (formerly Orolia Academic Partnership Program).

\bibliographystyle{IEEEtran}
    \bibliography{Bibliography/IEEEabrv, Bibliography/Articles, Bibliography/Books, Bibliography/InProceedings, Bibliography/Misc}

\end{document}